\documentclass[trackchanges, twocolumn, tighten]
{aastex701}

\usepackage{multirow}

\begin{document}

\title{FALCON (FERMI–ALMA Calibrator ObservatioN): Cross-matching ALMA Calibrators with Fermi-LAT Blazars}

\author[0009-0000-5913-8555]{Kanako Narita}
\affiliation{Department of Astronomy, Graduate School of Science, The University of Tokyo, 7-3-1 Hongo, Bunkyo-ku, Tokyo 133-0033, Japan}
\affiliation{National Astronomical Observatory of Japan, National Institutes of Natural Sciences, 2-21-1 Osawa, Mitaka, Tokyo 181-8588, Japan}
\email{naritakn@g.ecc.u-tokyo.ac.jp}

\author[0000-0002-7272-1136]{Yoshiyuki Inoue}
\affiliation{College of Systems Engineering and Science, Shibaura Institute of Technology, 307 Fukasaku, Minuma, Saitama, Saitama, 337-8570, Japan}
\affiliation{Interdisciplinary Theoretical \& Mathematical Science Center (iTHEMS), RIKEN, 2-1 Hirosawa, 351-0198, Japan}
\affiliation{Kavli Institute for the Physics and Mathematics of the Universe (WPI), UTIAS, The University of Tokyo, 5-1-5 Kashiwanoha, Kashiwa, Chiba 277-8583, Japan}
\email{naritakn@g.ecc.u-tokyo.ac.jp}

\author[0000-0001-6469-8725]{Bunyo Hatsukade}
\affiliation{National Astronomical Observatory of Japan, 2-21-1 Osawa, Mitaka, Tokyo 181-8588, Japan}
\affiliation{Graduate Institute for Advanced Studies, SOKENDAI, Osawa, Mitaka, Tokyo 181-8588, Japan}
\affiliation{Department of Astronomy, Graduate School of Science, The University of Tokyo, 7-3-1 Hongo, Bunkyo-ku, Tokyo 133-0033, Japan}
\email{naritakn@g.ecc.u-tokyo.ac.jp}

\author[0000-0002-0921-8837]{Yasushi Fukazawa}
\affiliation{Department of Physics, Graduate School of Advanced Science and Engineering, Hiroshima University, 1-3-1 Kagamiyama, Higashi-Hiroshima, Hiroshima 739-8526, Japan}
\email{fukazawa@astro.hiroshima-u.ac.jp}

\begin{abstract}
Blazars are active galactic nuclei with relativistic jets closely aligned with our line of sight, producing Doppler-boosted emission from radio to gamma rays. Millimeter emission near the synchrotron self-absorption turnover probes compact regions at the jet base, yet large statistical studies in this regime remain limited.
We construct the FALCON (FERMI--ALMA Calibrator ObservatioN) catalogue, a millimeter-selected blazar sample by cross-matching the ALMA Calibrator Source Catalogue with the Third Data Release of the Fourth LAT AGN Catalog (4LAC-DR3). 
After removing duplicates, we obtain 949 unique gamma-ray–associated sources, corresponding to approximately 25 \% of the 4LAC-DR3 blazar population.
For the subset with available redshift measurements, we compute millimeter and gamma-ray luminosities and find a strong, statistically significant correlation that remains robust after accounting for redshift effects. Fermi-detected blazars exhibit systematically flatter millimeter spectral indices than gamma-ray non-detected ALMA calibrators, while the redshift-controlled luminosity correlation becomes progressively weaker from high-synchrotron-peaked BL Lacs toward flat-spectrum  radio quasars. These results indicate an intrinsic connection between millimeter and gamma-ray emission, consistent with a common origin in relativistic jets governed by Doppler boosting and time-averaged jet power.

\end{abstract}



\section{Introduction} 
\label{sec:intro}

Blazars are radio-loud active galactic nuclei (AGN) whose relativistic jets are oriented close to the line of sight, so that their Doppler-boosted, non-thermal emission dominates the observed spectrum from radio to gamma rays \citep{1995PASP..107..803U}. Their broadband spectral energy distribution (SED) shows two characteristic humps: a low-energy component produced by synchrotron radiation from relativistic electrons in the jet, and a high-energy component, peaking in the gamma-ray band, generally attributed to inverse-Compton (IC) scattering. A central open question in jet physics is how these two components are physically connected, and why the gamma-ray output varies so widely across the population, with many radio-bright blazars remaining comparatively faint in gamma rays \citep{2011ApJ...742...27L,2012A&A...541A.160G}. The detectability of a blazar in gamma rays has been linked to a range of global jet properties, including faster jet speeds and brightness \citep{2009ApJ...696L..17K,2011A&A...535A..69N}, smaller viewing angles \citep{2009ApJ...696L..22L}, larger apparent jet opening angles \citep{2009A&A...507L..33P,2010A&A...519A..45O}, stronger variability \citep{2011ApJS..194...29R}, and the presence of multiple inverse-Compton components \citep{2010ApJ...716...30A,2012JPhCS.355a2033T}. Enhanced gamma-ray activity has also been associated with dynamical events within the jet, such as the ejection of superluminal components \citep{2011ApJ...726L..13A} and high-frequency radio flares \citep{2011A&A...532A.146L}. These diverse connections underscore the need to characterize the relationship between the synchrotron and IC components across a large and uniformly selected sample.

Among the accessible wavebands, the millimeter regime offers a particularly clean view of the innermost jet. At these frequencies the emission lies near the synchrotron self-absorption turnover and samples the optically thin synchrotron spectrum of the most compact regions at the jet base, where contamination from thermal emission associated with the accretion flow and its host galaxy and from extended radio lobes is minimal \citep{2009A&A...508..107G}. Millimeter measurements therefore probe the synchrotron emission from precisely the regions where particle acceleration is expected to feed the high-energy output, making the millimeter spectral properties of blazars a direct diagnostic of the physical conditions underlying their gamma-ray emission. Compact, parsec-scale radio cores of blazars have furthermore been statistically associated with the arrival directions of high-energy neutrinos \citep{2020ApJ...894..101P,2021A&A...650A..83H}. Because the millimeter band probes these same innermost jet regions, characterizing the millimeter properties of the blazar population would be also of multi-messenger interest.

A connection between millimeter and gamma-ray emission has been indicated by several previous studies. Using quasi-simultaneous \textit{Planck} and \textit{Fermi}-LAT data for $98$ blazars, \citet{2012ApJ...754...23L} reported a significant correlation between (sub-)millimeter and gamma-ray luminosities spanning several orders of magnitude, and found that gamma-ray--bright blazars tend to show flat or rising millimeter spectra, interpreted as a newly emerging, self-absorbed synchrotron component associated with activity in the inner jet. Cross-correlation analyses of radio-to-submillimeter and gamma-ray light curves for $54$ sources have likewise revealed correlated variability, with the bulk of the gamma-ray production located within or upstream of the millimeter core on (sub-)parsec scales \citep{2014MNRAS.441.1899F}. Detailed studies of individual sources of $one$ sources place the gamma-ray emission at parsec-scale distances along the jet, in some cases beyond the broad-line region \citep{2025A&A...704A.143R}, and resolved monitoring of objects such as the broad-line radio galaxy 3C~120 has linked enhanced millimeter activity to gamma-ray dissipation at sub-parsec distances from the central engine \citep{2015ApJ...799L..18T}. Together these results suggest that the disturbances in the inner jet responsible for millimeter flares are closely tied to episodes of enhanced gamma-ray emission.

Despite these advances, existing statistical samples remain limited. The correlation studies above rest on relatively small samples with a substantial fraction of upper limits, which restricts their statistical robustness and may bias the inferred relations. Wide-field surveys face complementary difficulties. The \textit{Planck} mission provided the first blind, all-sky samples selected at millimeter and submillimeter wavelengths \citep{2013A&A...550A.133P,2018A&A...619A..94P}, but its arcminute-scale positional uncertainties, photometric errors, and contamination from Galactic dust and dusty star-forming galaxies complicate the identification of the relatively rare blazars among far more numerous dusty star-forming galaxies, and constructing clean blazar samples from these data remains challenging even with dedicated mitigation \citep{2020A&A...644A..99P}. Deeper surveys with the South Pole Telescope and \textit{Herschel} reached fainter flux densities and recovered radio/millimeter--gamma correlations \citep{2022ApJ...939..117Z,2022MNRAS.513.6013M}, but cover limited sky areas ($\sim2500~\mathrm{deg}^2$ for SPT and $\sim642~\mathrm{deg}^2$ for Herschel) and therefore yield samples much smaller than the all-sky \textit{Fermi}-LAT survey. Previous statistical studies have thus generally been constrained by modest sample sizes and by selection effects such as flux limits and redshift dependence.

These limitations can be largely overcome using the ALMA Calibrator Source Catalogue \footnote{\url{https://almascience.nao.ac.jp/sc/}}. Over the past decade, the Atacama Large Millimeter/submillimeter Array (ALMA) has repeatedly observed thousands of compact, bright radio sources as calibration targets. These calibrators consist predominantly of flat-spectrum radio quasars (FSRQs) and BL~Lac objects and therefore form one of the largest and most homogeneous millimeter-selected samples of blazars currently available \citep{2018MNRAS.478.1512B}. Using the ALMA calibrator archive, \cite{2020A&A...641A..62P} demonstrated that the calibrator population is overwhelmingly dominated by blazars and constructed one of the largest millimeter-selected blazar samples through cross-matching with the Roma-BZCAT catalogue \citep{2015Ap&SS.357...75M}.
In fact, the ALMA Calibrator Source Catalogue contains several thousand compact radio sources observed across multiple millimeter bands, making it effectively the largest millimeter-selected blazar sample assembled to date. Owing to ALMA's sub-arcsecond angular resolution, source confusion and contamination from dusty star-forming galaxies are negligible compared with wide-beam surveys. Although calibration data have recently been exploited for blind continuum and spectral-line searches \citep[e.g.,][]{2024A&A...690A.258B}, and the calibrator population itself has been characterized at radio and millimeter wavelengths \citep{2024A&A...690A.258B, 2020A&A...641A..62P}, these studies have primarily focused on identifying and characterizing the blazar population contained in the ALMA calibrator archive. In contrast, its physical connection to gamma-ray emission has not yet been systematically investigated.

In this work, we construct the \textit{Fermi}--ALMA Calibrator
ObservatioN (FALCON) catalogue by cross-matching the ALMA Calibrator
Source Catalogue with the Third Data Release of the Fourth Catalog
of Active Galactic Nuclei Detected by the \textit{Fermi} Large Area
Telescope (4LAC-DR3; \citealt{2022ApJS..263...24A}). The FALCON
catalogue comprises all ALMA calibrators associated with 4LAC-DR3
sources, irrespective of redshift availability. Sources with reliable
redshift measurements constitute a subsample used for the
luminosity-based analyses. This large millimeter-selected,
gamma-ray-associated blazar sample enables the first systematic
investigation of the connection between millimeter and gamma-ray
emission in the ALMA calibrator population.

This paper is organized as follows. In Section~\ref{sec:data} we describe the ALMA and \textit{Fermi} data sets and the construction of the ALMA--\textit{Fermi} sample. In Section~\ref{sec:results} we present the observational results, including the matching statistics, redshift distributions, millimeter spectral properties, and gamma-ray luminosity distributions. In Section~\ref{sec:discussion} we investigate the millimeter--gamma-ray luminosity correlation and discuss its physical implications for particle acceleration and jet emission in blazars. Section~\ref{sec:summary} summarizes our conclusions. 
Throughout this paper, we adopt a standard $\Lambda$CDM cosmology with $H_0 = 70\ \mathrm{km\ s^{-1}\ Mpc^{-1}}$, $\Omega_{\Lambda} = 0.7$, and $\Omega_{\mathrm{M}} = 0.3$, broadly consistent with the Planck 2018 results \citep{2020A&A...641A...6P}.

\section{Data and Sample Construction} 
\label{sec:data}
\subsection{ALMA Calibrator Sample}
The ALMA Calibrator Source Catalogue is a comprehensive database of flux and structural measurements for compact calibrator sources. 
ALMA calibration sources are selected to be bright, compact, and largely unresolved radio sources with little extended structure and accurately known positions ($\lesssim0.003''$). 
The angular resolution varies between observations owing to changes in the ALMA array configuration and observing frequency. Since the calibrators are selected to be compact and are generally unresolved, configuration-dependent effects on their measured flux densities are expected to be small for most sources. However, sources with extended emission may be affected by differences in the spatial scales sampled by different configurations.
The catalogue includes sources south of declination $+45^\circ$ with typical flux densities above $\sim10$\,mJy at 90\,GHz, providing suitable calibrators for reliable amplitude, phase, bandpass, and pointing calibration.
As of 1~January~2021, it contains 9,274 individual ALMA measurements of more than 3,362 sources obtained across all observing bands.
For each source, the catalogue provides the flux density and its uncertainty at a given frequency, as well as information on angular structure (via $uv_{\rm min}$ and $uv_{\rm max}$ limits) and polarization for a subset of calibrators.

The catalogue aggregates ALMA monitoring data together with information from several external ``seed'' databases, including the Karl G. Jansky Very Large Array (VLA), the Submillimeter Array (SMA), the Australia Telescope Compact Array (ATCA), the Combined Radio All-Sky Targeted Eight GHz Survey (CRATES), and Very Long Baseline Interferometry (VLBI) reference frame catalogues (e.g., \citealt{2011MNRAS.414.2528P}).
Coordinates are tied to the VLBI reference frame.
For Bands 3, 6, and 7, the standard Time Division Mode (TDM)
setup comprises four spectral windows (SPWs), two in each
sideband, with an effective bandwidth of 1.875~GHz per SPW.
In Bands 6 and 7, the catalogue flux densities are typically
averaged over all four SPWs, giving a total effective bandwidth
of 7.5~GHz.
For Band 3, separate flux-density measurements are provided
for the lower and upper sidebands, each combining two SPWs
with an effective bandwidth of 3.75~GHz.
The reported frequency for each Band 3 measurement is the
midpoint between the central frequencies of the two SPWs.
The two sideband frequencies are separated by approximately
12~GHz; this separation is distinct from the effective
bandwidth of each measurement.


The absolute flux scale is referenced to solar-system primary calibrators, although the quoted flux-density uncertainties may not fully account for uncertainties in the planetary flux models. According to the ALMA Calibrator Source Catalogue documentation, the observation and reduction procedures changed in mid-2012, and measurements obtained before this change may have underestimated flux-density uncertainties.We therefore conservatively restricted our analysis to observations obtained on or after 2013 January 1.

The catalogue is publicly accessible via the ALMA Science Portal and supports queries by position, time, frequency, and flux range.
Further details of the calibration procedures and the use of these sources for amplitude and phase calibration are described in the ALMA Technical Handbook and in \citet{2014Msngr.155...19F}.

\subsection{Fermi-LAT Gamma-Ray Sample}
\label{sec:fermi_sample}

We use gamma-ray data from the fourth catalogue of AGN detected by the \textit{Fermi} Large Area Telescope (LAT) in its third data release \citep[4LAC-DR3;][]{2022ApJS..263...24A}, which is based on the 4FGL-DR3 source catalogue \citep{2022ApJS..260...53A}.
These catalogues provide spectral and flux information for sources detected over the energy range from 100~MeV to 100~GeV.
In this study, we specifically use the photon index and the energy flux integrated over the LAT energy range reported in the catalogue.
Redshift information is taken from the \textit{Fermi} catalogue when available and supplemented using the 5th edition of the Roma-BZCAT catalogue \citep{2015Ap&SS.357...75M}, as described in Section~\ref{sec:counterpart}.

\subsection{Identification of Gamma-ray Counterparts}
\label{sec:counterpart}
To identify gamma-ray counterparts with the ALMA calibrator sources, we followed the statistical approach adopted in optical--radio counterpart studies, such as the Blazar Radio and Optical Survey (BROS) sample \citep{2020ApJ...901....3I}. Rather than assigning counterparts solely based on nearest-neighbor distances, this approach characterizes the association through the angular separation distribution, allowing true associations to be statistically separated from chance coincidences.

For each ALMA catalogue entry, we considered all gamma-ray sources from the \textit{Fermi}-LAT 4LAC-DR3 catalogue \citep{2022ApJS..263...24A} located within a maximum angular separation of $3^\circ$. Using all ALMA--gamma-ray source pairs, we constructed the number counts of neighboring gamma-ray sources in bins of angular separation between $r$ and $r+dr$. The bin width was set to $dr = 0.001^\circ$ (3.6~arcsec), providing sufficient resolution to characterize the small-separation excess while remaining appropriate for the positional uncertainties of gamma-ray sources.

\begin{figure}
\epsscale{1.0}
\plotone{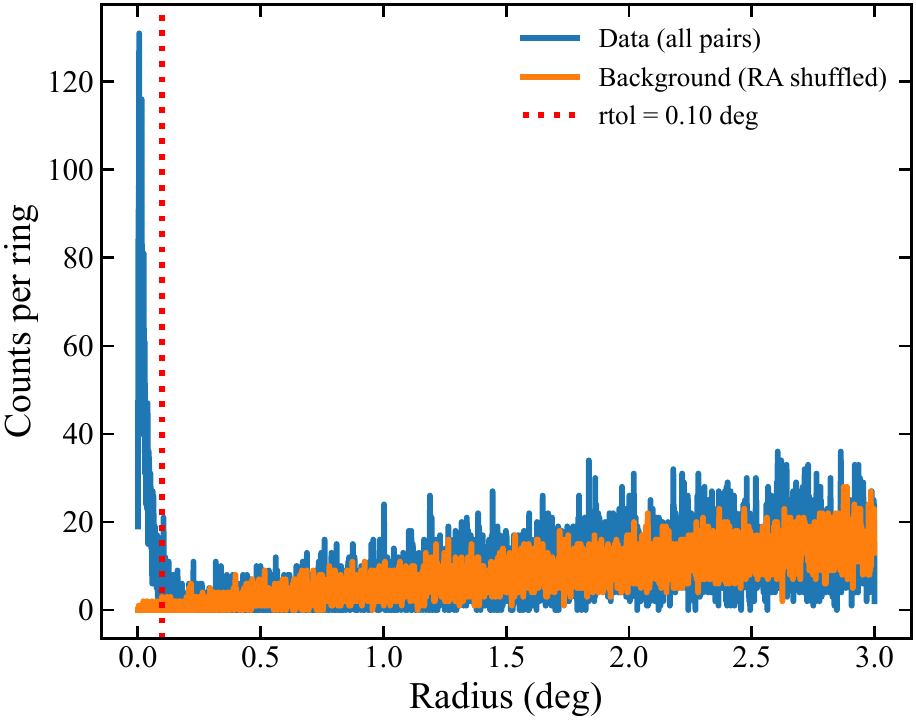}
\caption{Distance distribution between ALMA and Fermi sources. Orange line
indicates the expected number of association by chance coincidence and red
dashed line indicates $\mathrm{r}_{tol}$ = 0.1 degree position (see the text for details).
}
\label{fig:match}
\end{figure}

To estimate the contribution from chance coincidences, we generated a randomized catalogue by permuting the right ascension of the ALMA sources while preserving their declination distribution. The same counting procedure was applied to this randomized catalogue to derive the background distribution. For a random distribution of sources, the number of neighboring pairs is expected to increase approximately linearly with angular separation, and this component was treated as the background.
Figure~\ref{fig:match} shows the radial distributions of ALMA–Fermi source pairs together with the expected background estimated from the randomized catalogue. A clear excess of pairs is observed at small angular separations relative to the background, indicating the presence of genuine ALMA–Fermi associations.
By comparing the cumulative distributions of the observed and background counts, we defined the number of true associations within radius $r$ as
\begin{equation}
N_{\rm true}(<r) = N_{\rm obs}(<r) - N_{\rm bg}(<r),
\end{equation}
where $N_{\rm obs}$ and $N_{\rm bg}$ denote the cumulative observed and background pair counts, respectively. Because genuine counterparts contribute only at small angular separations, whereas chance coincidences accumulate with the surveyed area, the cumulative difference $N_{\rm true}(<r)$ rises steeply at small $r$ and then gradually saturates at larger radii.  We identify the total number of true associations, $N_{\rm true}(\infty)$, with this saturation value, attained at separations several times larger than $r_{\rm tol}$, beyond which the observed counts are statistically consistent with chance coincidence alone.

Following the BROS analysis, we quantified the quality of the matching using two metrics: the contamination fraction,
\begin{equation}
f_{\rm cont}(r) = \frac{N_{\rm bg}(<r)}{N_{\rm obs}(<r)},
\end{equation}
and the overlooked fraction,
\begin{equation}
f_{\rm over}(r) = \frac{N_{\rm true}(\infty) - N_{\rm true}(<r)}{N_{\rm true}(\infty)}.
\end{equation}
The tolerance radius $r_{\rm tol}$ was determined empirically by balancing these two quantities.

In this study, we adopted a tolerance radius of $r_{\rm tol} = 0.10^\circ$. 
At this radius, we estimate a contamination fraction of approximately 1\% and an overlooked fraction of approximately 17\%. These values are comparable to those reported in previous optical--radio counterpart studies, despite the substantially larger positional uncertainties of gamma-ray sources.

\begin{figure*}[t]
\centering
\includegraphics[width=\textwidth]{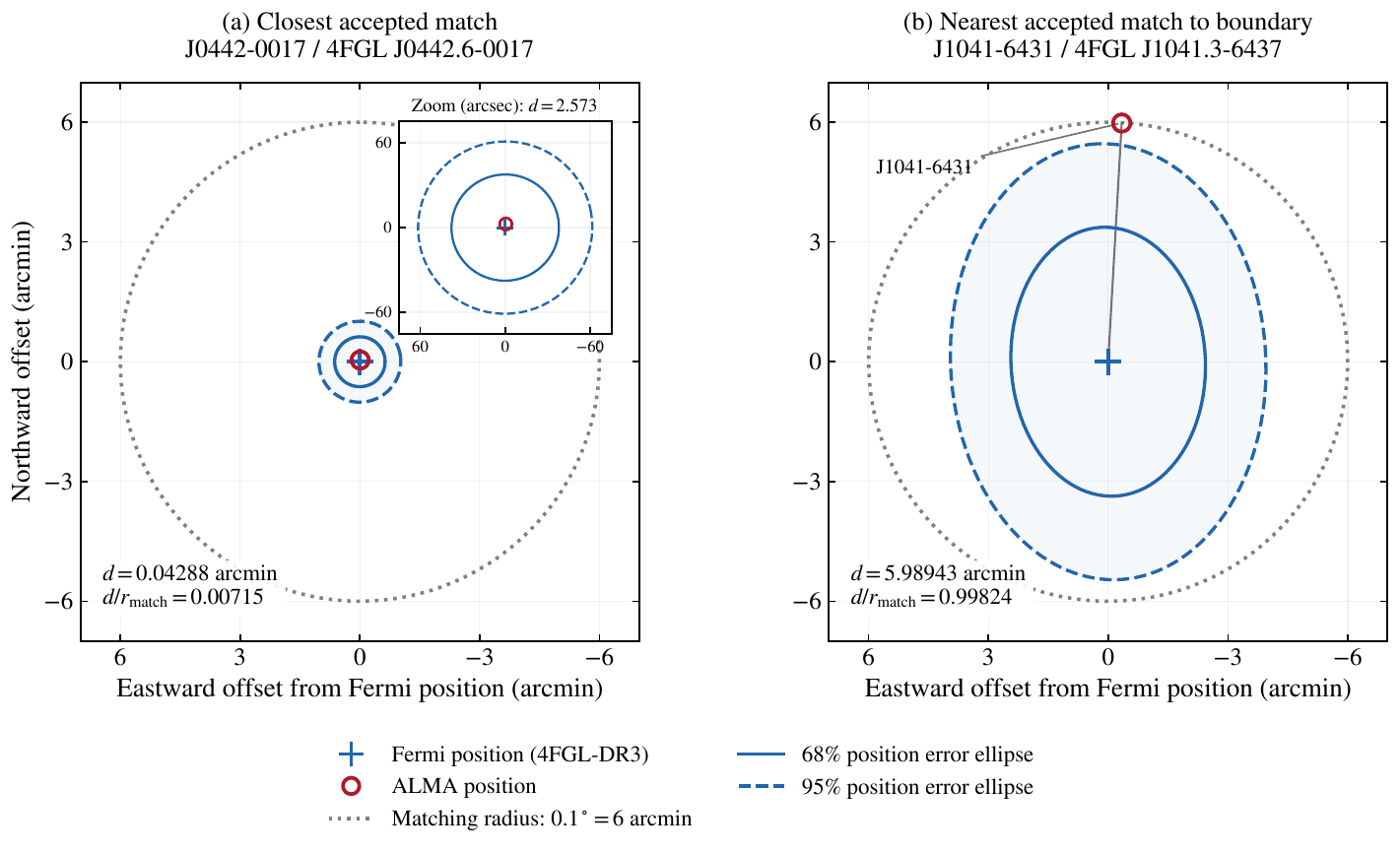}
\caption{
Examples of positional matching between ALMA calibrators
and \textit{Fermi}-LAT sources.
(a) The selected pair with the smallest angular separation:
J0442$-$0017 and 4FGL J0442.6$-$0017
($d=2.573$~arcsec).
(b) The selected pair closest to the adopted matching boundary:
J1041$-$6431 and 4FGL J1041.3$-$6437
($d=5.989$~arcmin).
Blue crosses and red open circles mark the \textit{Fermi}-LAT
and ALMA catalogue positions, respectively.
Gray dotted circles indicate the adopted matching radius
of $0.1^\circ$ (6~arcmin), rather than a localization
confidence region.
The quantity $r_{\rm match}$ used in the panels is identical
to $r_{\rm tol}$ defined in the text.
Blue solid and dashed ellipses show the 68\% and 95\%
positional uncertainty regions from 4FGL-DR3, respectively.
Both main panels use the same angular scale, with offsets
measured relative to the \textit{Fermi}-LAT position.
The inset in panel (a) shows an enlarged view in arcseconds.
The ALMA position in panel (b) lies within the adopted
matching radius but outside the 95\% positional
uncertainty ellipse.
}
\label{fig:matching_examples}
\end{figure*}

Figure~\ref{fig:matching_examples} illustrates the positional
correspondence between ALMA and \textit{Fermi}-LAT sources for
two representative pairs selected by our matching criterion:
the pair with the smallest angular separation,
J0442$-$0017 and 4FGL J0442.6$-$0017 (2.573~arcsec),
and the pair closest to the adopted matching boundary,
J1041$-$6431 and 4FGL J1041.3$-$6437 (5.989~arcmin).
Both main panels use the same angular scale, and an inset
shows an enlarged view of the closest pair.
The 68\% and 95\% positional uncertainty ellipses from
4FGL-DR3 are shown for comparison and were not used as
additional selection criteria.
Thus, the boundary pair satisfies the adopted
$r_{\rm tol}=0.10^\circ$ criterion even though its ALMA
position lies outside the 95\% positional uncertainty ellipse.

In three cases, multiple ALMA calibrators initially satisfied the angular-separation criterion for the same Fermi-LAT source. We resolved these ambiguous associations by comparing the ALMA positions with the published counterpart coordinates in 4LAC-DR3 and retaining, in each case, the unique ALMA calibrator located within $1''$ of the counterpart position. After resolving these multiple matches, the final Fermi-matched sample comprises 949 ALMA calibrators associated with 949 distinct Fermi-LAT sources.

To supplement the redshift information for this final sample,
we additionally cross-matched the Fermi-matched sources with the
fifth edition of the Roma-BZCAT catalogue
\citep{2015Ap&SS.357...75M}, using a matching radius of $3''$.
We chose Roma-BZCAT because it compiles blazar classifications
and literature-based redshifts, providing a useful supplementary
source of redshift information for our blazar-dominated sample.
It includes BL Lac objects and candidates (BZB), flat-spectrum
radio quasars (BZQ), sources with a substantial host-galaxy
contribution to their spectral energy distributions (BZG), and
blazars of uncertain type (BZU). When a reliable redshift was
unavailable in 4LAC-DR3, the corresponding value from Roma-BZCAT
was adopted.

\section{Results}
\label{sec:results}

\subsection{Matching Results}
We cross-matched the ALMA Calibrator Source Catalogue with the
Fermi-LAT 4LAC-DR3 catalogue based on positional coincidence, and
summarize the results in Table~\ref{tab:match_stats_band}.
The number of unique ALMA calibrators with observations obtained
on or after 2013 January 1 is 3288, 1288, and 2531 in Bands~3, 6,
and 7, respectively. Among these, 940 (29\%), 627 (48\%), and
837 (33\%) sources have adopted Fermi-LAT counterparts.
Because many calibrators were observed in more than one
ALMA band, these band-specific samples overlap. After combining
the three band-specific samples and removing duplicates across
bands, we obtain 949 unique \textit{Fermi}-LAT-associated ALMA
calibrators. These 949 sources constitute the full FALCON catalogue,
irrespective of redshift availability.

Among the FALCON sources, 645, 470, and 597 objects in Bands~3, 6, and 7, respectively, have reliable redshift measurements. These counts precede the flux-validity and luminosity cuts used to define the final luminosity-analysis samples. Sources without reliable redshift measurements remain part of the FALCON catalogue but are excluded from analyses
requiring luminosity measurements.


\begin{table*}
\centering
\caption{Summary of the parent ALMA calibrator samples and their
cross-matching statistics with the \textit{Fermi}-LAT 4LAC-DR3
catalogue. Counts are based on unique ALMA sources within each band
with observations obtained on or after 2013 January 1.
Parent counts are measured before resolving multiple ALMA
candidates associated with the same \textit{Fermi}-LAT source;
matched counts include only the adopted counterparts.
Because some sources were observed in multiple ALMA bands,
the band-specific samples overlap.
After removing duplicates across bands, the full FALCON catalogue
contains 949 unique \textit{Fermi}-LAT-associated ALMA calibrators,
irrespective of redshift availability.
Redshifts are taken from 4LAC-DR3 and supplemented with
Roma-BZCAT when a valid 4LAC-DR3 redshift is unavailable.
These counts precede the flux-validity and luminosity cuts
used to define the luminosity-analysis samples.}
\label{tab:match_stats_band}
\begin{tabular}{lccc}
\hline\hline
Band &
Parent ALMA calibrators &
4LAC-matched ALMA calibrators &
Matched sources with redshift \\
\hline
Band 3 & 3288 & 940 & 645 \\
Band 6 & 1288 & 627 & 470 \\
Band 7 & 2531 & 837 & 597 \\
\hline
\end{tabular}
\end{table*}

\subsection{Redshift Distributions}

\begin{figure*}
\epsscale{0.8}
\plotone{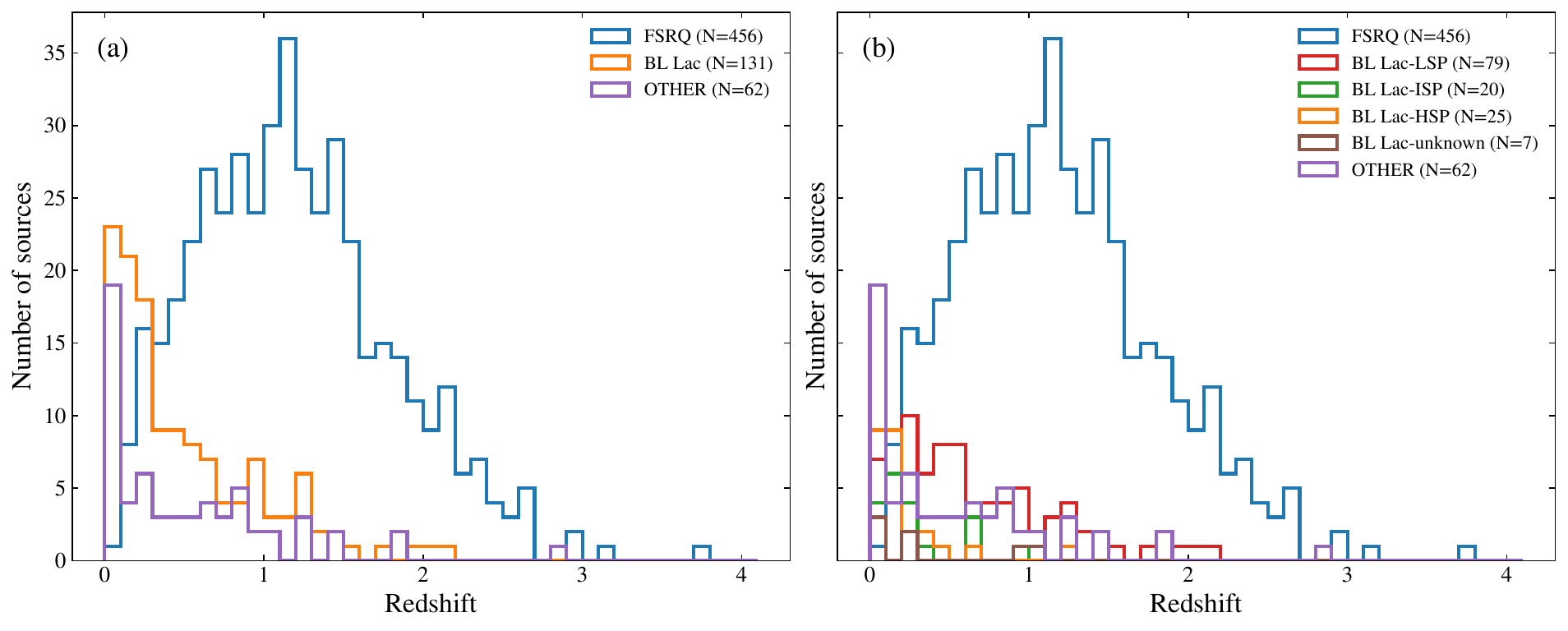}
\caption{
Redshift distributions of the Fermi-detected sources used in this study.
(a) Redshift distributions by broad source class (FSRQ, BL Lac, and other AGNs).
(b) Redshift distributions by blazar subclass (FSRQ, LSP, ISP, and HSP).
FSRQs extend systematically to higher redshifts, whereas BL Lac objects,
particularly HSPs, are preferentially found at lower redshifts.
}
\label{fig:fermi_detected_by_class}
\end{figure*}

Figure~\ref{fig:fermi_detected_by_class} shows the redshift
distributions of the 649 unique FALCON sources included in
at least one of the Band 3, 6, or 7 luminosity-analysis samples,
requiring $L_{\rm mm}\geq10^{40}\ {\rm erg\,s^{-1}}$
in the corresponding band.
Each source is counted once, and both panels use the same
source list.

This luminosity-selected FALCON subset reproduces the
well-established qualitative difference between FSRQs and
BL~Lac objects, with FSRQs extending to higher redshifts
and BL~Lac objects concentrated at lower redshifts.
The broad redshift coverage, particularly for FSRQs, highlights the importance of controlling for redshift effects in luminosity-based correlation analyses, motivating the use of partial correlation methods in the following sections.

\begin{figure}
\epsscale{0.8}
\plotone{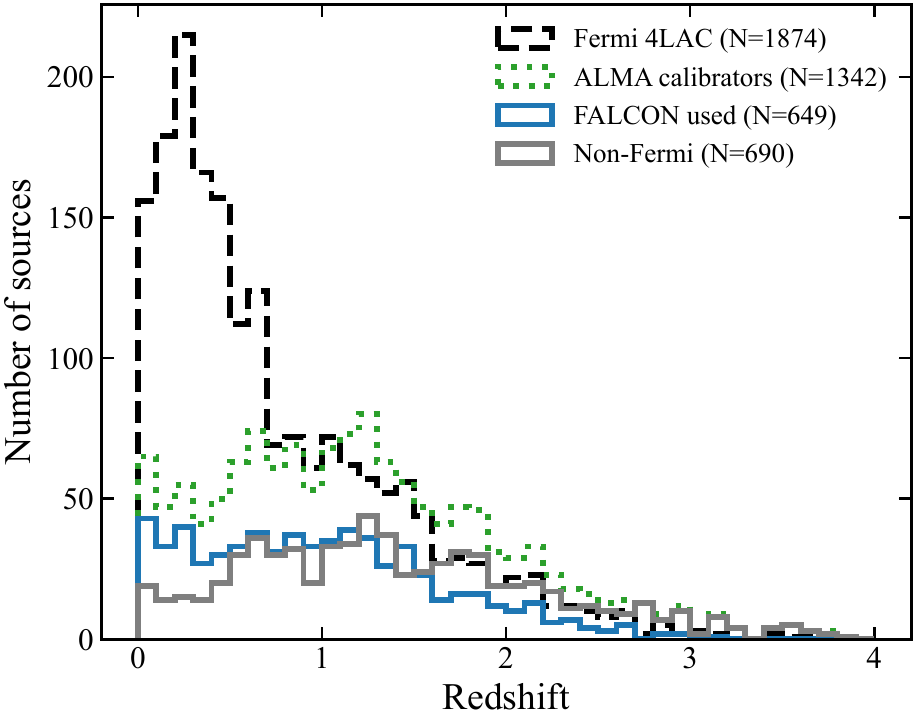}
\caption{Redshift distributions of the 4LAC-DR3 blazar sample (black dashed), ALMA calibrators with available redshift measurements (green dotted), Fermi-matched ALMA calibrators (blue), and ALMA calibrators without a Fermi-LAT counterpart (gray). The figure compares the redshift distribution of the FALCON sample with that of the parent 4LAC-DR3 blazar population, as well as with ALMA calibrators lacking a Fermi-LAT counterpart.
}
\label{fig:redshift_known_dist}
\end{figure}

Figure~\ref{fig:redshift_known_dist} compares the redshift distributions of the parent 4LAC-DR3 blazar sample, ALMA calibrators with available redshift measurements, Fermi-matched ALMA calibrators, and ALMA calibrators without a Fermi-LAT counterpart. Compared with the parent 4LAC-DR3 sample, the FALCON sample contains relatively fewer low-redshift blazars, indicating that not all nearby gamma-ray blazars are included in the ALMA calibrator sample. The Fermi-detected and non-detected ALMA calibrators largely overlap in redshift. Although non-detected sources exhibit a slightly broader tail toward higher redshifts, both populations span a similar redshift range. 
This demonstrates that gamma-ray non-detection cannot be attributed solely to distance effects, indicating that additional physical and observational factors contribute to the observed difference between the two populations.

\subsection{Gamma-ray photon index distributions}
\label{sec:photon_index}
\begin{figure*}
\epsscale{0.8}
\plotone{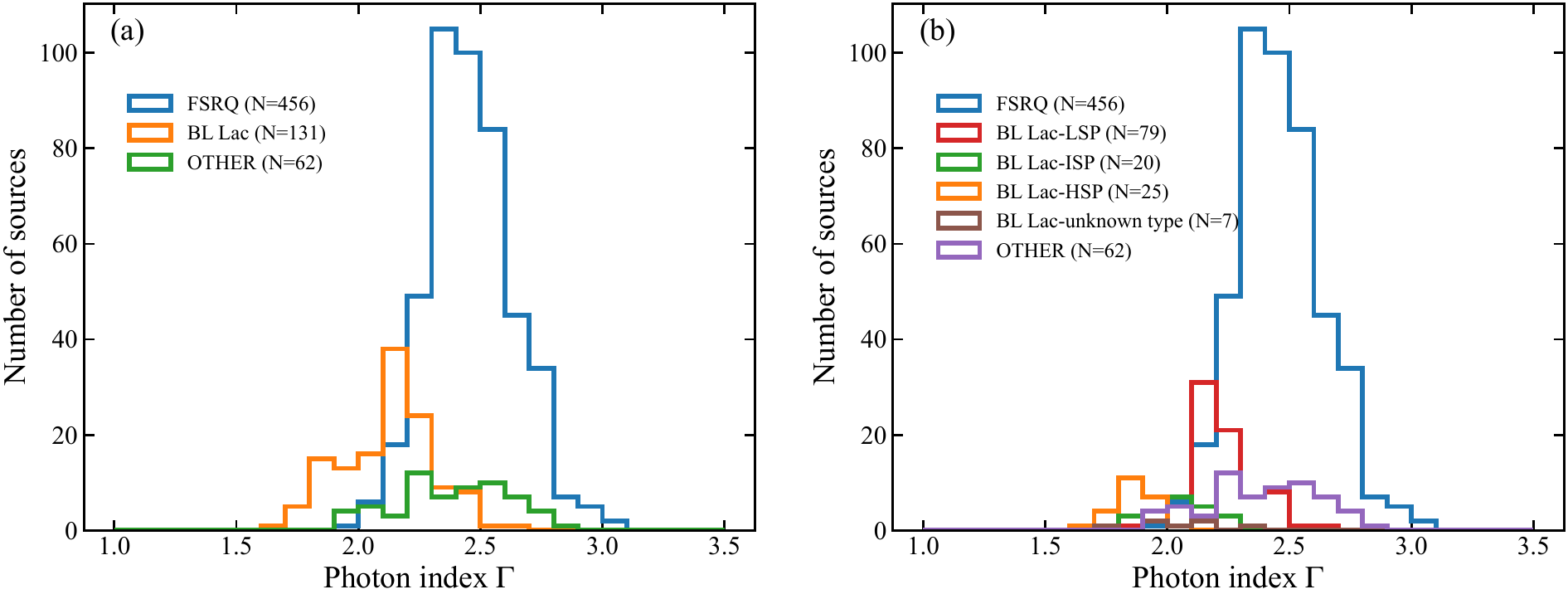}
\caption{
Photon-index distributions of the Fermi-LAT detected sources used in this study.
(a) Broad source classes (FSRQ, BL Lac, and other AGNs).
(b) Blazar subclasses (FSRQ, LSP, ISP, and HSP).
FSRQs exhibit systematically softer gamma-ray spectra than BL Lac objects, while the photon index becomes progressively harder from LSP to ISP and HSP BL Lac objects.
}
\label{fig:photon_index}
\end{figure*}

Figure~\ref{fig:photon_index} shows the distributions of the Fermi-LAT photon index ($\Gamma_{\rm LAT}$) for the sources used in this study. FSRQs exhibit systematically softer gamma-ray spectra than BL Lac objects, with median photon indices of 2.44 and 2.14, respectively. When the BL Lac population is further divided into SED subclasses based on the synchrotron peak frequency (low-, intermediate-, and high-synchrotron-peaked; LSP, ISP, and HSP), the median photon index decreases systematically from LSP ($\Gamma_{\rm LAT, med}=2.19$) to ISP ($\Gamma_{\rm LAT, med}=2.09$) and HSP ($\Gamma_{\rm LAT, med}=1.86$), demonstrating progressively harder gamma-ray spectra toward higher synchrotron-peak frequencies. This trend is consistent with the well-established spectral sequence of blazars \citep{1998MNRAS.299..433F,1998MNRAS.301..451G,2017MNRAS.469..255G}

\subsection{Spectral Index Distributions}
\label{sec:sp_index}
For sources observed in both ALMA Band 3 and Band 6, we compute the two-point spectral index defined as

\begin{equation}
\alpha_{36} =
\frac{\log(S_6 / S_3)}{\log(\nu_6 / \nu_3)},
\end{equation}

\begin{figure}
\epsscale{0.8}
\plotone{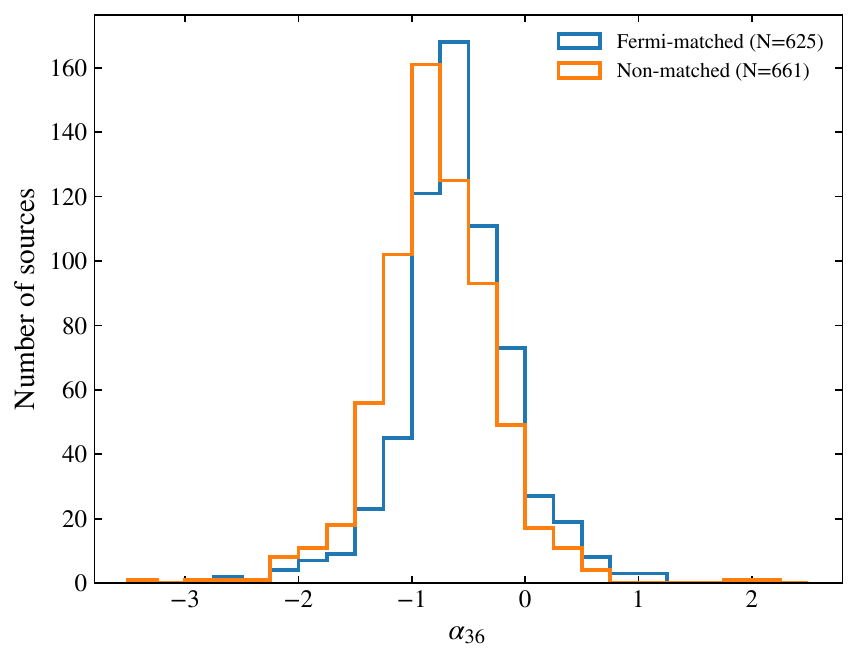}
\caption{Distribution of the two-point millimeter spectral index $\alpha_{36}$ for sources observed in both ALMA Band 3 and Band 6. Blue and orange histograms represent Fermi-detected and non-detected sources, respectively.
}
\label{fig:alpha36}
\end{figure}

where $S_3$ and $S_6$ are the median flux densities within each band and $\nu_3$ and $\nu_6$ are the corresponding median observing frequencies. 
Median flux densities are adopted to reduce the impact of variability across multiple observing epochs.
Figure~\ref{fig:alpha36} shows the distribution of $\alpha_{36}$ for Fermi-detected and non-detected sources. 
The median spectral index for Fermi-detected sources is $-0.60$, whereas non-Fermi sources exhibit a steeper median value of $-0.80$. 
A Kolmogorov–Smirnov test yields a probability of $p\approx10^{-12}$, demonstrating that the two distributions are statistically distinct.

Despite partial overlap between the two populations, the systematic flattening observed in gamma-ray–detected sources indicates that millimeter spectral slopes are closely connected to the physical conditions responsible for high-energy emission.

\begin{figure}
\epsscale{0.8}
\plotone{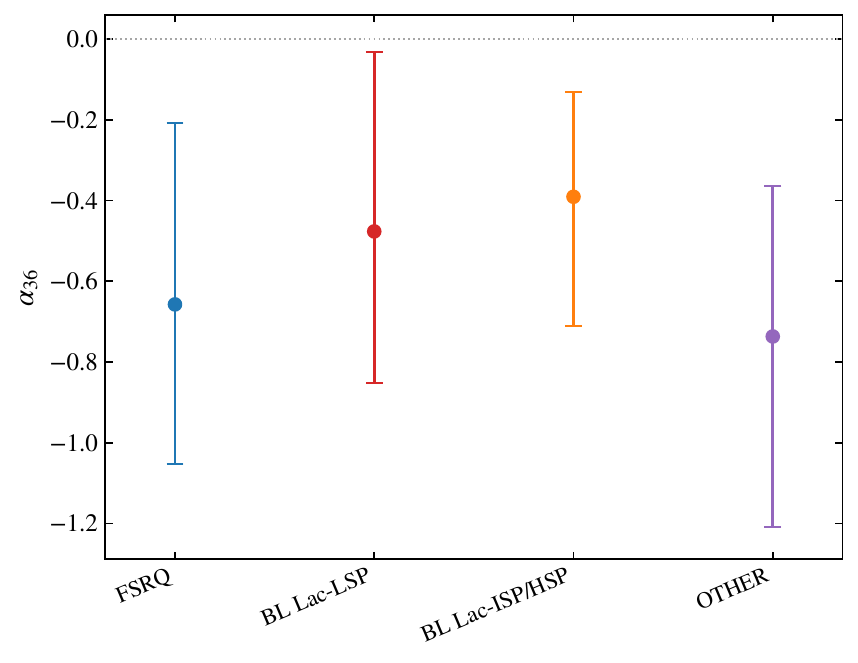}
\caption{Median \(\alpha_{36}\) values for the Fermi-matched sources used in the
luminosity analysis. Error bars indicate the 16th--84th percentile
ranges.
}
\label{fig:alpha36_subclass}
\end{figure}

We further examined the distribution of $\alpha_{36}$ within the Fermi-detected sample by dividing the sources into blazar subclasses (Figure~\ref{fig:alpha36_subclass}). The plotted points indicate the median spectral index for each subclass, while the error bars represent the 16th--84th percentile ranges. 
FSRQs and LSP BL Lac objects exhibit steeper median millimeter spectra, with $\widetilde{\alpha}_{36}=-0.643_{-0.363}^{+0.393}$ and $-0.476_{-0.334}^{+0.408}$, respectively, whereas the combined ISP/HSP BL Lac sample shows a flatter median value of $\widetilde{\alpha}_{36}=-0.403_{-0.261}^{+0.222}$. The quoted lower and upper uncertainties correspond to the 16th and 84th percentiles, respectively.
The distributions of FSRQs and LSP BL Lac objects are also broader, with standard deviations of 0.459 and 0.435, respectively, compared with 0.316 for the ISP/HSP BL Lac sample. Although formal statistical tests do not provide strong evidence for differences in the distribution widths, the systematic flattening toward ISP/HSP BL Lac objects is qualitatively consistent with expectations from the blazar sequence and motivates the physical interpretation discussed in Section~\ref{sec:discussion_subclass}.

\subsection{Luminosity Calculation}

To investigate the intrinsic connection between millimeter and gamma-ray emission, we computed rest-frame luminosities for both bands using standard cosmological relations. 
For each source with known redshift, the millimeter luminosity was characterized using the band-integrated quantity $\nu L_\nu$, defined as
\begin{equation}
    \nu L_{\nu} = 4\pi D_L^2\, \nu_{\rm obs}\, S_{\nu_{\rm obs}}\,
    (1+z)^{\alpha_{\rm mm}-1},
\end{equation}
where $D_L$ is the luminosity distance, $S_{\nu_{\rm obs}}$ is the observed flux density at frequency $\nu_{\rm obs}$, and $\alpha_{\rm mm}$ is the millimeter spectral index defined by $S_\nu \propto \nu^{-\alpha_{\rm mm}}$. The factor $(1+z)^{\alpha_{\rm mm}-1}$ accounts for the $K$-correction of the observed millimeter flux density to the rest frame.
This definition enables a direct comparison of energy output across different frequencies while minimizing assumptions about the detailed shape of the millimeter spectrum.
For the $K$-correction, we used the millimeter spectral index derived from the Band~3--6 spectral index distribution introduced in \ref{sec:sp_index}. 
Because $\alpha_{36}$ was defined as $S_\nu \propto \nu^{\alpha_{36}}$, whereas the present $K$-correction assumes $S_\nu \propto \nu^{-\alpha_{\rm mm}}$, we adopted $\alpha_{\rm mm} = -\alpha_{36}$ for sources with measured Band~3--6 spectral indices. 
For sources lacking a valid flux-density measurement in either Band 3 or Band 6, we adopt $\alpha_{\rm mm}=0.60$, the rounded median of the measured indices for 625 Fermi-matched sources.
Replacing this fallback value with the 16th and 84th percentiles
of the measured distribution (0.139 and 0.979) leaves sample
membership unchanged and changes the full-sample partial
correlation coefficients by less than 0.013 in all three bands.

We computed the isotropic gamma-ray luminosity assuming a power-law photon spectrum with photon index $\Gamma$. 
The observed energy flux $S_{\rm obs}$ was converted into luminosity as
\begin{equation}
L_{\gamma} = 4\pi D_L^2\, S_{\rm obs}\, (1+z)^{\Gamma-2},
\end{equation}
where $D_L$ is the luminosity distance and the factor $(1+z)^{\Gamma-2}$ represents the $K$-correction appropriate for a power-law spectrum.

The resulting luminosities span $\sim10^{41}$--$10^{47}$~erg~s$^{-1}$ at millimeter wavelengths and $\sim10^{42}$--$10^{49}$~erg~s$^{-1}$ in gamma rays, comparable to those of known blazars detected by \textit{Fermi}-LAT.
These values form the basis for the correlation analysis discussed in Section~\ref{sec:correlation_analysis}.


\begin{table*}
\centering
\caption{Number of sources in the final luminosity--luminosity
analysis samples after requiring valid positive flux densities
and redshifts and applying the criterion
$L_{\rm mm}\geq10^{40}\ {\rm erg\ s^{-1}}$.}
\label{tab:used_sample_band}
\begin{tabular}{lccccccc}
\hline\hline
Band &
Total used &
FSRQ &
BL Lac LSP &
BL Lac ISP &
BL Lac HSP &
BL Lac unk. &
Other AGNs \\
\hline
Band 3 & 642 & 451 & 77 & 20 & 25 & 7 & 62 \\
Band 6 & 469 & 345 & 61 & 9  & 10 & 3 & 41 \\
Band 7 & 593 & 421 & 75 & 15 & 21 & 7 & 54 \\
\hline
\end{tabular}
\end{table*}

\subsection{Gamma-Ray Luminosity Distributions}
\begin{figure}
\epsscale{0.8}
\plotone{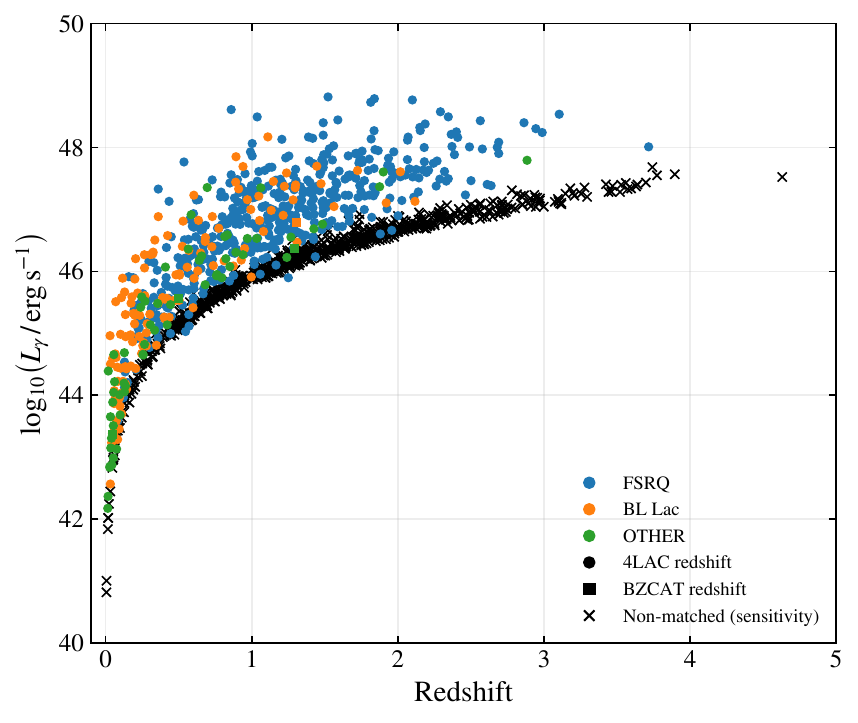}
\caption{Gamma-ray luminosity as a function of redshift for Fermi-detected blazars, together with upper limits for non-detections.
Upper limits are derived from the Fermi--LAT energy flux sensitivity assuming a power-law spectrum.
}
\label{fig:gamma_lumi_distribution}
\end{figure}

Figure \ref{fig:gamma_lumi_distribution} compares the gamma-ray luminosity--redshift distribution of Fermi-detected blazars with upper limits for non-detections, clearly illustrating the flux-limited nature of the sample and the strong redshift-dependent detection boundary.
While the non-detected sources do not form a clearly distinct population in the \(L_\gamma\)--\(z\) plane, it remains possible that they preferentially probe blazars with weaker relativistic beaming.
In this interpretation, the lower gamma-ray luminosities primarily reflect differences in Doppler boosting rather than intrinsic power.
In contrast, the radio emission may be less strongly affected by beaming if it includes a significant contribution from downstream jet regions with more isotropic emission.
Such wavelength-dependent beaming effects could account for the observed trends without invoking a fundamentally distinct source class.

\subsection{Catalogue Description}
\label{sec:catalog}

\begin{deluxetable*}{lll}
\tablecaption{FALCON Catalogue Description\label{tab:catalog}}
\tablehead{
\colhead{Column Name} & \colhead{Unit} & \colhead{Description}
}
\startdata
SOURCE\_ID
    & \nodata
    & IAU-style source identifier derived from the ALMA position \\
ALMA\_NAME
    & \nodata
    & Source aliases listed in the ALMA calibrator catalogue \\
RA
    & deg
    & Right ascension of the ALMA calibrator in the ICRS/J2000 frame \\
DEC
    & deg
    & Declination of the ALMA calibrator in the ICRS/J2000 frame \\
REDSHIFT
    & \nodata
    & Adopted redshift \\
Z\_ORIGIN
    & \nodata
    & Origin of the adopted redshift (4LAC-DR3, Roma-BZCAT, or unknown) \\
S\_B3
    & Jy
    & Median ALMA Band~3 flux density \\
E\_S\_B3
    & Jy
    & Median quoted uncertainty of the ALMA Band~3 flux density \\
S\_B6
    & Jy
    & Median ALMA Band~6 flux density \\
E\_S\_B6
    & Jy
    & Median quoted uncertainty of the ALMA Band~6 flux density \\
S\_B7
    & Jy
    & Median ALMA Band~7 flux density \\
E\_S\_B7
    & Jy
    & Median quoted uncertainty of the ALMA Band~7 flux density \\
FGL\_NAME
    & \nodata
    & Associated 4LAC-DR3/Fermi-LAT source name \\
GAMMA\_FLUX
    & erg\,cm$^{-2}$\,s$^{-1}$
    & Fermi-LAT gamma-ray energy flux in the 0.1--100\,GeV band \\
E\_GAMMA\_FLUX
    & erg\,cm$^{-2}$\,s$^{-1}$
    & Uncertainty of the gamma-ray energy flux \\
PHOTON\_INDEX
    & \nodata
    & Fermi-LAT power-law photon index \\
E\_PHOTON\_INDEX
    & \nodata
    & Uncertainty of the power-law photon index \\
ALPHA36
    & \nodata
    & Two-point millimeter spectral index derived from ALMA Bands~3 and 6 \\
CLASS
    & \nodata
    & Source classification (FSRQ, BL Lac, or BCU) \\
SED\_TYPE
    & \nodata
    & Spectral energy distribution classification (LSP, ISP, or HSP) \\
SEP\_FERMI
    & deg
    & Angular separation between the ALMA and Fermi-LAT positions \\
\enddata

\tablecomments{
The FALCON catalogue contains the 949 ALMA calibrators associated with
Fermi-LAT sources in 4LAC-DR3.
Redshifts are adopted preferentially from 4LAC-DR3 and supplemented with
Roma-BZCAT where available.
Sources without reliable redshift measurements remain in the catalogue.
ALMA flux densities and their quoted uncertainties are the median values
of all available measurements in each observing band.
The millimeter spectral index is defined as
$\alpha_{36}=\log(S_{6}/S_{3})/\log(\nu_{6}/\nu_{3})$.
The full catalogue is available in FITS format.\\
ALMA Calibrator Source Catalogue:
\url{https://almascience.nao.ac.jp/alma-data/documents-and-tools/cycle10/flux-service-of-the-alma-source-catalogue}.\\
Fourth LAT AGN Catalog (4LAC): \citet{2020ApJ...892..105A}.
}
\end{deluxetable*}

\begin{deluxetable*}{lrrrlrrrrlll}
\tablecaption{Sample of the FALCON Catalogue\label{tab:falcon_sample}}
\tablehead{
\colhead{Source ID} &
\colhead{R.A.} &
\colhead{Decl.} &
\colhead{$z$} &
\colhead{$z$ origin} &
\colhead{$S_{\rm B3}$} &
\colhead{$S_{\rm B6}$} &
\colhead{$S_{\rm B7}$} &
\colhead{$\alpha_{36}$} &
\colhead{4FGL name} &
\colhead{Class} &
\colhead{SED}
}
\startdata
J000118.02-074626.9 & 0.32510 & -7.77415 & \nodata & unknown & 0.109 & 0.075 & 0.083 & $-0.461$ & 4FGL J0001.2-0747 & BL Lac & LSP \\
J000211.98-215309.9 & 0.54992 & -21.88607 & \nodata & unknown & 0.388 & \nodata & 0.168 & \nodata & 4FGL J0001.8-2153 & BCU & LSP \\
J000404.91-114858.4 & 1.02048 & -11.81622 & 1.310 & 4LAC-DR3 & 0.286 & 0.248 & 0.137 & $-0.176$ & 4FGL J0003.9-1149 & BL Lac & LSP \\
J000435.66-473619.6 & 1.14856 & -47.60545 & 0.880 & 4LAC-DR3 & 0.394 & \nodata & 0.190 & \nodata & 4FGL J0004.4-4737 & FSRQ & LSP \\
J000557.18+382015.1 & 1.48823 & 38.33754 & 0.229 & 4LAC-DR3 & 0.876 & \nodata & 0.413 & \nodata & 4FGL J0005.9+3824 & FSRQ & LSP \\
J000613.89-062335.3 & 1.55789 & -6.39315 & 0.347 & 4LAC-DR3 & 1.750 & 0.910 & 0.690 & $-0.806$ & 4FGL J0006.3-0620 & BL Lac & LSP \\
J004204.55+232001.1 & 10.51894 & 23.33363 & 1.426 & 4LAC-DR3 & 0.307 & \nodata & 0.135 & \nodata & 4FGL J0042.2+2319 & FSRQ & LSP \\
J064007.18-125315.4 & 100.02993 & -12.88761 & 0.135 & 4LAC-DR3 & 0.039 & \nodata & 0.025 & \nodata & 4FGL J0640.0-1253 & BL Lac & HSP \\
J064151.13-032048.6 & 100.46305 & -3.34683 & 1.196 & 4LAC-DR3 & 0.518 & 0.361 & 0.161 & $-0.445$ & 4FGL J0641.7-0320 & FSRQ & LSP \\
J064326.45+085738.0 & 100.86019 & 8.96056 & 0.882 & 4LAC-DR3 & 0.434 & 0.205 & 0.220 & $-0.924$ & 4FGL J0643.3+0857 & FSRQ & LSP \\
\enddata

\tablecomments{
R.A. and Decl. are the ALMA positions in the ICRS/J2000 frame and are
given in degrees.
The ALMA Band~3, Band~6, and Band~7 flux densities are median values
and are given in Jy.
The two-point millimeter spectral index is defined as
$\alpha_{36}=\log(S_{\rm B6}/S_{\rm B3})/
\log(\nu_{\rm B6}/\nu_{\rm B3})$.
The redshift origin identifies the catalogue from which the adopted
redshift was obtained.
An entry of ``unknown'' indicates that no reliable redshift is available.
This table presents the first 10 entries in the FITS catalogue; the full
catalogue of 949 Fermi-LAT-associated ALMA calibrators is available in
FITS format.
}
\end{deluxetable*}

The FALCON catalogue comprises 949 unique ALMA calibrators
associated with sources in the \textit{Fermi}-LAT 4LAC-DR3 catalogue,
irrespective of redshift availability. An association with a
4LAC-DR3 source is required for inclusion in FALCON, whereas ALMA
calibrators without a \textit{Fermi}-LAT counterpart are not included.

\begin{figure*}
\epsscale{1.0}
\plotone{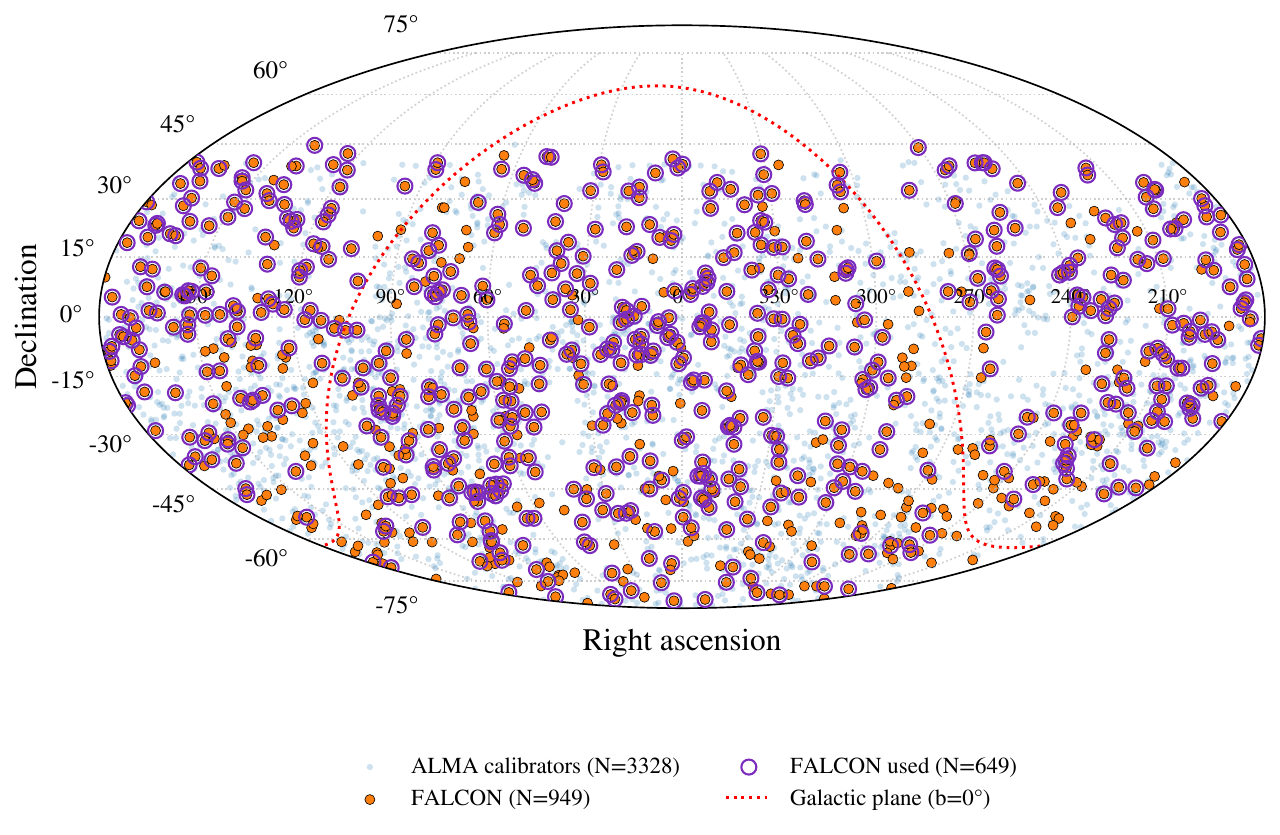}
\caption{Sky distribution of ALMA calibrators in equatorial coordinates. Light blue points show all ALMA calibrators, orange points indicate the FALCON sources, and purple open circles denote the subset with available redshift measurements used for the luminosity analysis. The red dotted curve marks the Galactic plane ($b=0^\circ$). The ALMA footprint reflects the sky accessible from the ALMA site, while the lower density of FALCON sources toward the Galactic plane is caused by reduced Fermi-LAT sensitivity to extragalactic sources in regions of strong diffuse Galactic gamma-ray emission.
}
\label{fig:sky_dist}
\end{figure*}

A description of the FALCON catalogue is provided in
Table~\ref{tab:catalog}, and a representative sample of the catalogue
is presented in Table~\ref{tab:falcon_sample}.
The catalogue contains the source identifier, ALMA position,
adopted redshift and its origin, ALMA flux densities and uncertainties
in Bands~3, 6, and 7, the associated \textit{Fermi}-LAT source name,
gamma-ray energy flux and its uncertainty, photon index and its uncertainty, the two-point millimeter spectral
index ($\alpha_{36}$) derived from the ALMA Band~3 and Band~6 flux
densities, blazar classification, SED type, and the angular separation
between the ALMA and \textit{Fermi}-LAT positions. Redshifts are taken
preferentially from the 4LAC-DR3 catalogue and supplemented with the
Roma-BZCAT where available, as described in Section~2.2. Sources
without reliable redshift measurements remain part of the FALCON
catalogue but are excluded from analyses requiring luminosity
measurements. Flux densities correspond to the median ALMA
measurements in each observing band.
The catalogue is available at \url{https://github.com/naritakn/FALCON-Survey-FERMI-ALMA-Calibrator-ObservatioN}.
Figure \ref{fig:sky_dist} shows the sky distribution of FALCON sources. As expected, the ALMA calibrators are confined to the sky accessible from the ALMA site, with a sharp boundary near declination $\delta \lesssim 45^\circ$. 
The resulting anisotropy therefore reflects the pointing distribution of ALMA calibration observations rather than any intrinsic property of the blazar population.
A reduced density of Fermi--LAT associations is also visible toward the Galactic plane, particularly in the Galactic center region, where the strong diffuse Galactic gamma-ray emission increases the detection threshold and decreases the sensitivity to extragalactic point sources.

\section{Discussion}
\label{sec:discussion}
\subsection{Millimeter Spectral Indices and the Blazar Sequence}
The flatter and narrower $\alpha_{36}$ distribution of the
ISP/HSP BL Lac sample is qualitatively consistent with
the blazar sequence, in which the synchrotron peak frequency systematically increases from FSRQs through LSP and ISP to HSP BL Lac objects.
Previous studies have shown that the overall millimeter/submillimeter spectral shape is related to the activity state of relativistic jets and the occurrence of gamma-ray flares (e.g., \citealt{2012ApJ...754...23L}). \cite{2016A&A...596A.106P} also investigated radio-to-submillimeter spectra across different AGN subclasses, including BL Lac objects and quasars. Building on these studies, our larger ALMA–Fermi sample enables a statistical examination of the millimeter spectral slope across blazar subclasses, including its dependence on synchrotron peak frequency.
The systematic flattening
of $\alpha_{36}$ from FSRQs toward ISP/HSP BL Lac objects therefore
suggests that the low-frequency synchrotron continuum itself also
reflects the underlying SED sequence. These results suggest that the
blazar sequence is manifested not only in the synchrotron peak
frequency but also in the millimeter spectral slope.
Consequently, ALMA Band 3 and Band 6 probe different parts of the synchrotron spectrum among these subclasses.
In this picture, the millimeter
emission of FSRQs and LSP BL Lacs is more sensitive to source-to-source
differences in synchrotron peak location and/or variability, leading to
steeper and broader \(\alpha_{36}\) distributions. By contrast, ISP/HSP
BL Lacs may more often probe a relatively stable low-energy tail or
flat-spectrum core component. Given the lack of strong statistical
evidence for different distribution widths, however, we regard this as a
suggestive trend rather than a conclusive result.

\subsection{Correlation between $L_{\rm mm}$ and $L_\gamma$}
\label{sec:correlation_analysis}
\subsubsection{Global correlation}
\label{sec:correlation_global}
For the luminosity--luminosity analysis, we excluded sources with
$L_{\rm mm}<10^{40}\ {\rm erg\ s^{-1}}$. These sources exhibit
unusually low millimeter luminosities compared with the main ALMA
calibrator population and are therefore treated as outliers in the
present analysis. 
After requiring valid positive flux densities and redshifts
and applying this luminosity cut, the final samples consist
of 642, 469, and 593 sources in Bands~3, 6, and 7, respectively,
as summarized in Table~\ref{tab:used_sample_band}.
This corresponds to an increase by a factor of $\sim$5--6 compared to previous
studies based on samples of $\sim$100 blazars \citep{2012ApJ...754...23L, 2022ApJ...939..117Z, 2022MNRAS.513.6013M}, which were often limited by small
sample sizes and a substantial fraction of upper limits.
The significantly larger sample enables a more robust statistical
characterization of the millimeter--gamma-ray connection, including reliable
partial correlation measurements and class-dependent analyses.

\begin{figure*}
\gridline{
  \fig{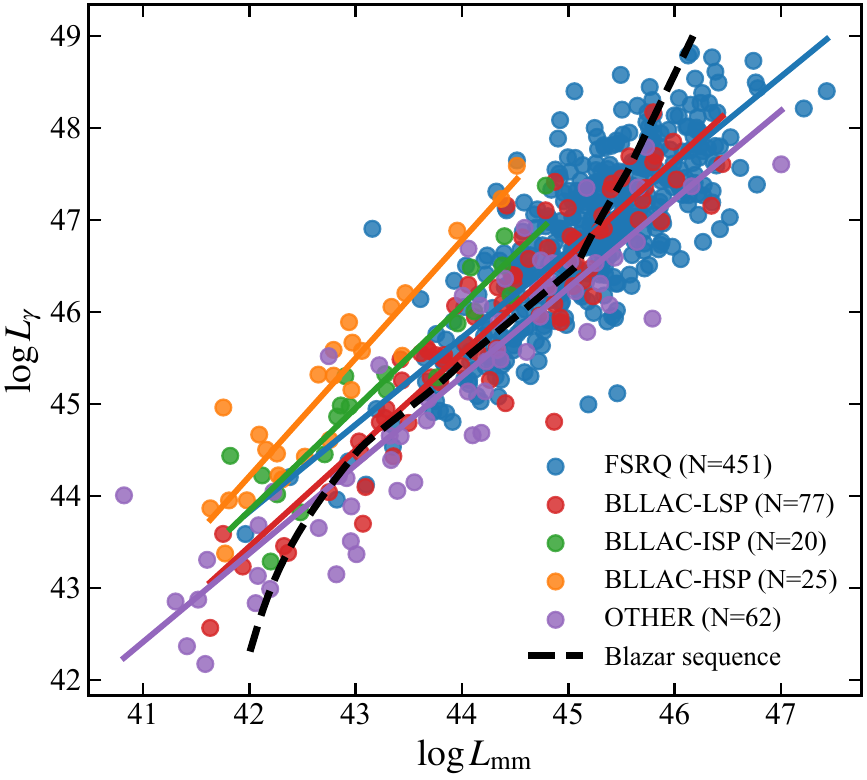}{0.33\textwidth}{(a) Band 3}
  \fig{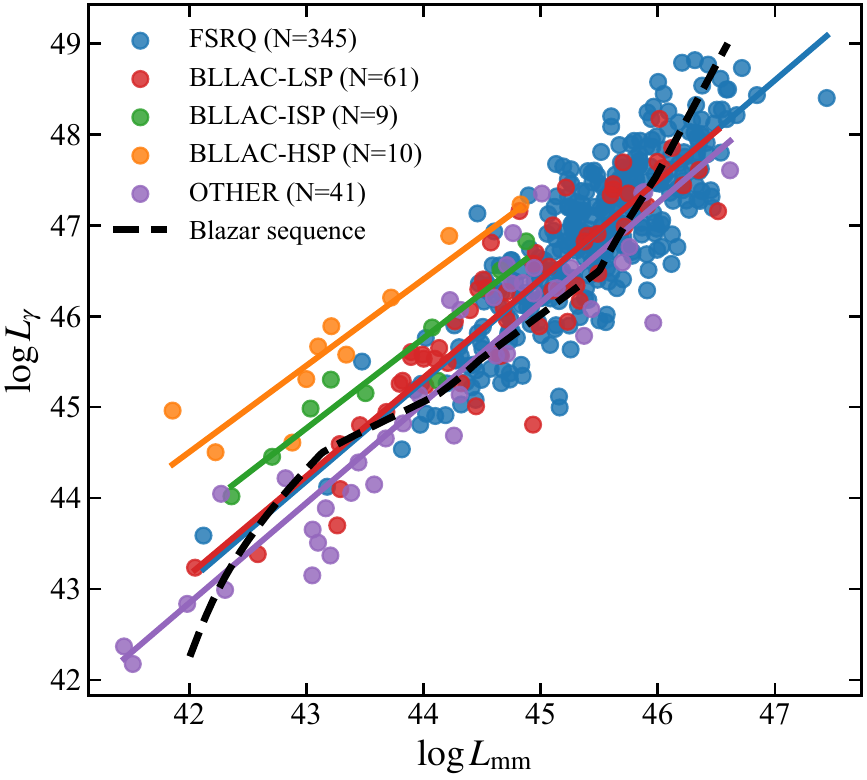}{0.33\textwidth}{(b) Band 6}
  \fig{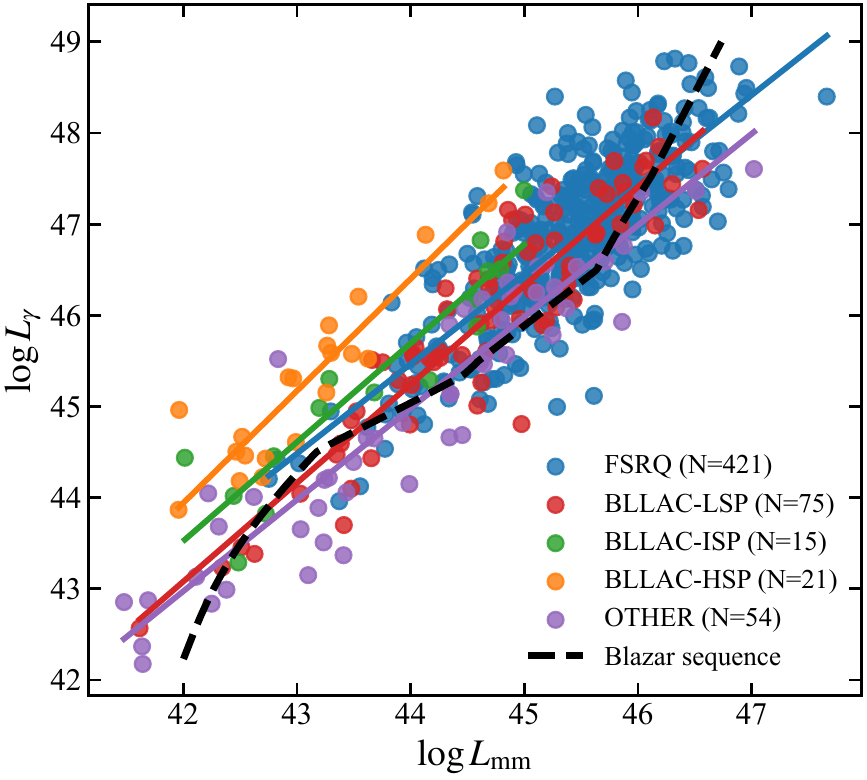}{0.33\textwidth}{(c) Band 7}
}
\caption{Correlation between the millimeter luminosity $L_{\rm mm}$
and the gamma-ray luminosity $L_{\gamma}$ for ALMA Band~3 (left),
Band~6 (center), and Band~7 (right) sources. Different colors denote
blazar subclasses (FSRQs, BL~Lac objects, and other AGNs). Solid lines
indicate the best-fit relations for each subclass, while the dashed
curve represents the blazar ``main sequence'' from
\citet{2017MNRAS.469..255G}.}
\label{fig:plot_lmm_lgamma}
\end{figure*}

We first computed the Spearman rank-order correlation coefficients for each
pair of millimeter luminosity, gamma-ray luminosity, and redshift. To 
account for the strong common dependence of luminosities on redshift in
flux-limited samples, we performed a partial correlation analysis following
\citet{1992A&A...256..399P}, removing the redshift dependence from both
luminosities. The significance of the partial correlation coefficients was evaluated assuming a $t$-distribution. The correlation coefficients and the best-fit regression parameters for each observing band and source class are summarized in Table \ref{tab:fisher_corr}.
As shown in Figure~\ref{fig:plot_lmm_lgamma} for Bands~3, 6, and 7,
we find a strong and statistically significant correlation between 
millimeter and gamma-ray luminosities in all ALMA bands, with 
Spearman rank coefficients of $\rho \sim 0.8$--0.9.
The correlation remains significant even after controlling for redshift,
with partial correlation coefficients of $\sim 0.5$--0.6. The slope of the
luminosity--luminosity relation is close to unity 
($L_{\gamma} \propto L_{\rm mm}^{\,\sim 1}$) across all bands, indicating
a nearly linear scaling between the two emission components.
The dashed curve in Figure~\ref{fig:plot_lmm_lgamma} schematically indicates the
classical blazar sequence proposed by \citet{2017MNRAS.469..255G}.
Although individual sources exhibit substantial scatter, the overall distribution broadly follows the expected trend. In particular, HSP BL Lac objects tend to lie systematically above FSRQs at a given millimeter luminosity, implying a larger gamma-ray-to-millimeter luminosity ratio.

\subsubsection{Dependence on Blazar Subclass}
\label{sec:discussion_subclass}
\begin{figure*}[t]
    \centering
    \includegraphics[width=\textwidth]{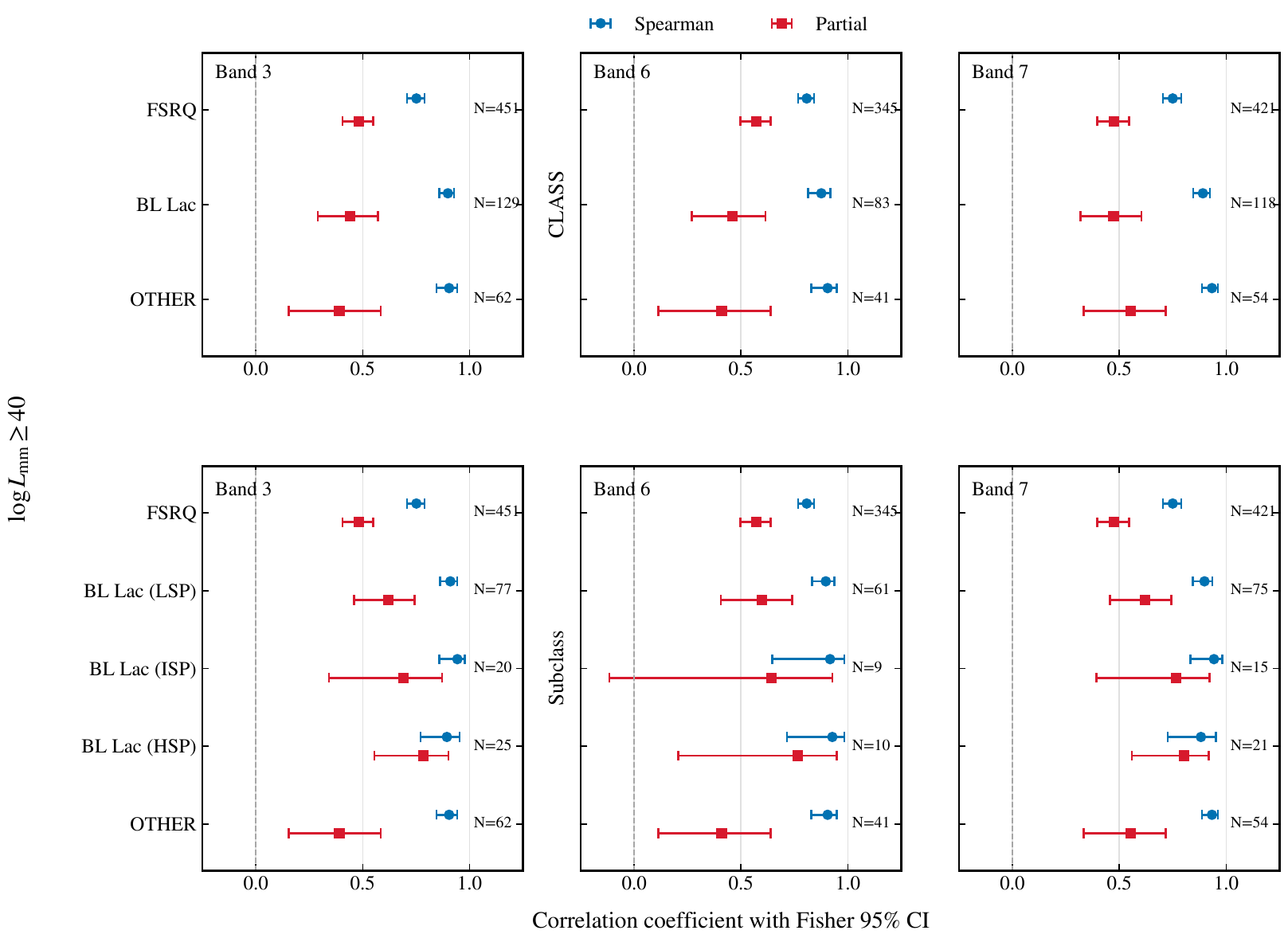}
    \caption{
    Correlation coefficients between millimeter and gamma-ray
    luminosities for sources with $\log L_{\rm mm}\geq40$,
    where $L_{\rm mm}$ is expressed in ${\rm erg\ s^{-1}}$.
    Columns correspond to ALMA Bands 3, 6, and 7, from left to right.
    Blue circles show the Spearman rank correlation coefficients,
    while red squares show the partial rank correlation coefficients
    after controlling for redshift.
    Horizontal error bars indicate approximate Fisher-transformed
    95\% confidence intervals.
    The upper panels show the results for FSRQs, BL Lacs,
    and other AGNs.
    The lower panels separate BL Lacs into LSP, ISP, and HSP
    subclasses, with FSRQs and other AGNs repeated for comparison.
    BL Lacs with unknown SED subclass are included in the
    upper-panel BL Lac samples but are not shown separately
    in the lower panels.
    The sample size for each group is indicated on the right
    of each panel.
    }
    \label{fig:fisher_correlation}
\end{figure*}

Figure~\ref{fig:fisher_correlation} shows the partial correlation coefficients for each blazar subclass together with their 95\% confidence intervals estimated using Fisher's $z$-transformation. Although the redshift-controlled correlation appears to become progressively weaker from HSP BL~Lacs toward FSRQs, the confidence intervals overlap substantially, especially for the ISP and HSP subclasses where the sample sizes are limited. Therefore, the observed subclass dependence should be regarded as suggestive rather than statistically conclusive.

This trend can be understood in the framework of the blazar sequence. In
HSP BL~Lacs, the synchrotron peak is located at X-ray energies and the
inverse-Compton peak at TeV energies. In this case, the millimeter and GeV
bands both sample the low-energy tails of the synchrotron and synchrotron self-Compton (SSC) components,
respectively. Since the synchrotron photons and inverse-Compton photons are
produced by the same underlying relativistic electron population in the SSC
scenario \citep[e.g.,][]{2010MNRAS.401.1570T, 2016ApJ...828...13I}, the millimeter and GeV gamma-ray luminosities are expected to be
closely linked. This provides a natural explanation for the relatively strong
partial correlation between $L_{\rm mm}$ and $L_{\gamma}$ observed in
HSP BL~Lacs.
This situation differs from that of FSRQs, in which the GeV
gamma-ray emission is generally interpreted as being dominated
by external Compton (EC) scattering of photons supplied by the
broad-line region and/or the dusty torus rather than by
SSC emission \citep[e.g.,][]{2016ApJ...830...94F}.
FSRQs also typically exhibit synchrotron peaks in the infrared and inverse-Compton peaks around MeV energies, placing the ALMA Band 3 emission much closer to the synchrotron peak than in HSP blazars. In this case, the millimeter
emission traces electrons radiating near the synchrotron peak, whereas the
Fermi-LAT GeV band is more likely to sample the high-energy side of the
inverse-Compton component. Moreover, in the EC scenario, the gamma-ray
luminosity depends not only on the jet electron population but also on the
strength and geometry of the external photon fields, such as radiation from
the broad-line region or dusty torus. These additional dependencies can weaken
the one-to-one correspondence between the millimeter and gamma-ray
luminosities, leading to a lower partial correlation toward FSRQs. This picture may also explain the class-dependent behavior of
$L_{\gamma}/L_{\rm mm}$, as FSRQs can exhibit enhanced millimeter emission
near the synchrotron peak without a proportional increase in the GeV gamma-ray luminosity.

This interpretation is further supported by the distributions of the Fermi-LAT photon index shown in Figure~\ref{fig:photon_index}. The median Fermi-LAT photon index  increases systematically from HSP ($\Gamma_{\rm LAT, med}=1.86$) to ISP ($\Gamma_{\rm LAT, med}=2.09$), LSP ($\Gamma_{\rm LAT, med}=2.19$), and FSRQs ($\Gamma_{\rm LAT, med}=2.44$), indicating progressively softer gamma-ray spectra along the blazar sequence. This systematic softening is consistent with the Fermi-LAT band sampling progressively higher-energy portions of the inverse-Compton component from HSP BL Lac objects to FSRQs, providing independent observational support for the SED-based interpretation presented above.

However, as shown in Figure~\ref{fig:fermi_detected_by_class}(b), the redshift distributions vary among the blazar subclasses, with HSP BL Lac objects preferentially found at lower redshifts, while FSRQs extend to systematically higher redshifts. Although the partial correlation analysis removes the first-order dependence on redshift, residual selection effects associated with the different redshift and luminosity coverage of each subclass cannot be completely excluded. Therefore, the subclass-dependent trends discussed above should be interpreted with appropriate caution.
This picture is also consistent with the distribution of sources in Figure~\ref{fig:plot_lmm_lgamma}. As noted in Section~\ref{sec:correlation_global}, HSP BL Lac objects lie systematically above FSRQs at a given millimeter luminosity, indicating a larger gamma-ray-to-millimeter luminosity ratio. This behavior is naturally explained by the harder gamma-ray spectra and higher synchrotron peak frequencies of HSP blazars (Figure~\ref{fig:photon_index}), whereas FSRQs typically have softer gamma-ray spectra and stronger Compton dominance. The broad scatter around the schematic blazar-sequence relation likely reflects variations in Doppler boosting, jet power, and source variability.

\subsubsection{Effects of Doppler Boosting and Jet Power}
The observed luminosities of relativistic jets are strongly modified by
Doppler boosting. For a single emission zone moving with bulk Lorentz factor
$\Gamma$ and viewing angle $\theta$, the Doppler factor is 
$\delta = [\Gamma(1-\beta\cos\theta)]^{-1}$. The bolometric synchrotron and SSC luminosities both scale as 
$L_\mathrm{bol} \propto \delta^{4}$, while the bolometric external Compton (EC)
luminosity scales more steeply, as $L_\mathrm{bol} \propto \delta^{6}$
\citep[e.g.,][]{1995ApJ...446L..63D,1996MNRAS.280...67G}. In our analysis 
we compare monochromatic luminosities $\nu L_{\nu}$ measured at fixed 
observed frequencies rather than bolometric quantities, so an additional 
dependence on the local spectral index of the emission enters the boost 
factor. Defining $\alpha_\mathrm{Syn}$ and $\alpha_\mathrm{IC}$ as the spectral 
indices ($S_{\nu} \propto \nu^{-\alpha}$) of the synchrotron and 
inverse-Compton components, the monochromatic luminosities at fixed 
observed frequency satisfy
\begin{equation}
\nu L_{\nu}^\mathrm{Syn,obs} \propto \delta^{3+\alpha_\mathrm{Syn}}, \quad
\nu L_{\nu}^\mathrm{SSC,obs} \propto \delta^{3+\alpha_\mathrm{IC}}, \quad
\nu L_{\nu}^\mathrm{EC,obs}  \propto \delta^{4+2\alpha_\mathrm{IC}}.
\end{equation}

These exponents imply that EC gamma-ray emission carries a 
substantially stronger Doppler dependence than synchrotron millimeter 
emission. Therefore, if source-to-source variations in the observed 
luminosities were dominated by variations in $\delta$ alone, one would 
generally expect a slope significantly steeper than unity in the 
$L_\mathrm{mm}$--$L_{\gamma}$ plane. The fact that the observed slope is consistent with unity, together 
with the residual scatter of $\sim 0.5$~dex after the partial correlation 
analysis, suggests that Doppler boosting is unlikely to be the sole 
driver of the observed correlation. Instead, both luminosities may 
primarily trace a common intrinsic property of the source, plausibly 
the time-averaged jet kinetic power $P_\mathrm{jet}$, while source-to-source 
differences in $\delta$, variability, radiative efficiency, and SED shape 
contribute to the remaining scatter. In this picture, $L_\mathrm{mm}$ and 
$L_{\gamma}$ are coupled through the energetics of the underlying 
relativistic electron population accelerated within the jet, with the 
millimeter synchrotron and Fermi-LAT inverse-Compton emission probing 
related, though not necessarily identical, parts of the jet emission 
region.

\subsubsection{Robustness against Non-simultaneous Observations}
A particularly notable feature of the sample is that the millimeter
flux densities and the Fermi-LAT energy fluxes are not measured 
simultaneously. The ALMA calibrator fluxes are derived from monitoring 
observations spanning many years (Section~\ref{sec:catalog}), and we adopt 
the median value at each band, effectively averaging over various individual states. The Fermi-LAT energy fluxes, in turn, are integrated over 
the entire DR3 observing interval. Despite this lack of simultaneity, and
despite the fact that blazars are known to vary by factors of several at 
both millimeter and gamma-ray wavelengths on timescales of weeks to 
years, the $L_\mathrm{mm}$--$L_{\gamma}$ correlation persists with high 
statistical significance.
This robustness against temporal averaging strengthens the interpretation 
that the correlation reflects an intrinsic, long-term property of the jet 
rather than coordinated short-term flaring activity. A similar argument was
made at lower radio frequencies by \citet{2011ApJ...741...30A} and at
submillimeter wavelengths by \citet{2012ApJ...754...23L} on the basis of 
quasi-simultaneous \textit{Planck}--Fermi data. Our results extend this 
conclusion to a substantially larger and millimeter-selected sample and
demonstrate that $L_\mathrm{mm}$ provides a useful, time-stable proxy for the
underlying jet power that drives the high-energy emission.

The present analysis is restricted to Fermi-LAT detected ALMA calibrators. Since the ALMA calibrator archive is not a uniform flux-limited survey, but rather a collection of sources selected for observational calibration purposes, extending the analysis to include gamma-ray non-detected sources requires careful consideration of the ALMA calibrator selection function and the inherent inhomogeneity of the sample. Such an analysis is beyond the scope of the present work and is left for future study.

\begin{table*}[tbh]
\centering
\caption{
Correlation analysis for sources with $\log L_{\rm mm}\geq40$,
where $L_{\rm mm}$ is in ${\rm erg\ s^{-1}}$.
The uncertainties on $\rho_{\rm S}$ and $\rho_{\rm partial}$
indicate approximate Fisher-transformed 95\% confidence intervals.
Partial correlations control for redshift.
The fits have the form
$\log L_{\gamma}=a\log L_{\rm mm}+b$,
with luminosities in ${\rm erg\ s^{-1}}$;
scatter is the standard deviation of the vertical residuals in dex.
The ALL rows also include BL Lacs with unknown SED subclass.
}
\label{tab:fisher_corr}
\begin{tabular}{llcccccc}
\hline\hline
Band & Class &
$N$ &
$\rho_{\rm S}$ &
$\rho_{\rm partial}$ &
Slope &
Intercept &
Scatter \\
\hline
\multirow{6}{*}{Band 3}
& FSRQ         & 451 & $0.75^{+0.04}_{-0.04}$ & $0.48^{+0.07}_{-0.07}$ & 0.95 &  4.13 & 0.54 \\
& BL Lac (LSP) &  77 & $0.91^{+0.03}_{-0.05}$ & $0.62^{+0.12}_{-0.16}$ & 1.05 & -0.67 & 0.47 \\
& BL Lac (ISP) &  20 & $0.94^{+0.03}_{-0.08}$ & $0.69^{+0.18}_{-0.35}$ & 1.11 & -2.96 & 0.37 \\
& BL Lac (HSP) &  25 & $0.89^{+0.06}_{-0.12}$ & $0.78^{+0.12}_{-0.23}$ & 1.28 & -9.72 & 0.37 \\
& OTHER        &  62 & $0.90^{+0.04}_{-0.06}$ & $0.39^{+0.19}_{-0.24}$ & 0.96 &  2.96 & 0.61 \\
& ALL          & 642 & $0.84^{+0.02}_{-0.02}$ & $0.48^{+0.06}_{-0.06}$ & 0.95 &  3.84 & 0.56 \\
\hline
\multirow{6}{*}{Band 6}
& FSRQ         & 345 & $0.81^{+0.03}_{-0.04}$ & $0.57^{+0.07}_{-0.08}$ & 1.10 & -3.20 & 0.48 \\
& BL Lac (LSP) &  61 & $0.90^{+0.04}_{-0.06}$ & $0.60^{+0.14}_{-0.19}$ & 1.08 & -2.35 & 0.45 \\
& BL Lac (ISP) &   9 & $0.92^{+0.07}_{-0.27}$ & $0.64^{+0.29}_{-0.76}$ & 0.99 &  2.11 & 0.25 \\
& BL Lac (HSP) &  10 & $0.93^{+0.06}_{-0.21}$ & $0.77^{+0.18}_{-0.56}$ & 0.95 &  4.56 & 0.34 \\
& OTHER        &  41 & $0.91^{+0.04}_{-0.08}$ & $0.41^{+0.23}_{-0.30}$ & 1.10 & -3.37 & 0.54 \\
& ALL          & 469 & $0.85^{+0.02}_{-0.03}$ & $0.55^{+0.06}_{-0.07}$ & 1.04 & -0.16 & 0.52 \\
\hline
\multirow{6}{*}{Band 7}
& FSRQ         & 421 & $0.75^{+0.04}_{-0.05}$ & $0.47^{+0.07}_{-0.08}$ & 0.98 &  2.23 & 0.54 \\
& BL Lac (LSP) &  75 & $0.90^{+0.04}_{-0.05}$ & $0.62^{+0.12}_{-0.16}$ & 1.08 & -2.35 & 0.47 \\
& BL Lac (ISP) &  15 & $0.94^{+0.04}_{-0.11}$ & $0.76^{+0.16}_{-0.37}$ & 1.08 & -2.01 & 0.45 \\
& BL Lac (HSP) &  21 & $0.88^{+0.07}_{-0.16}$ & $0.80^{+0.12}_{-0.24}$ & 1.23 & -7.69 & 0.37 \\
& OTHER        &  54 & $0.93^{+0.03}_{-0.05}$ & $0.55^{+0.16}_{-0.22}$ & 1.00 &  0.84 & 0.53 \\
& ALL          & 593 & $0.83^{+0.02}_{-0.03}$ & $0.48^{+0.06}_{-0.06}$ & 0.99 &  1.85 & 0.57 \\
\hline
\end{tabular}
\end{table*}

\section{Summary and Conclusions}
\label{sec:summary}

We present the results of the FALCON Survey, a systematic investigation of the millimeter--gamma-ray connection in blazars based on a large sample of ALMA calibrators cross-matched with Fermi-LAT sources. Our primary findings are as follows:

\begin{enumerate}

\item
We constructed the largest millimeter-wave blazar sample to date by cross-matching the ALMA Calibrator Source Catalogue with the Fermi-LAT 4LAC-DR3 catalogue.

The resulting FALCON catalogue contains 949 unique
\textit{Fermi}-associated ALMA calibrators observed across
ALMA Bands~3, 6 and 7, irrespective of redshift availability,
corresponding to approximately 25\% of the blazars and other
AGNs listed in 4LAC-DR3.

\item
We find that Fermi-detected sources exhibit systematically flatter millimeter
spectral indices compared to non-detected sources. The median spectral index is
$\alpha_{36} \sim -0.60$ for Fermi-detected sources and $\sim -0.80$ for non-detected
sources, with the difference being statistically significant. This result indicates
that millimeter spectral properties are closely connected to the physical conditions
responsible for gamma-ray emission.
Within the Fermi-detected sample, ISP/HSP BL~Lac objects exhibit systematically flatter millimeter spectra than FSRQs and LSP BL~Lac objects, qualitatively consistent with expectations from the blazar sequence.

\item
Using this sample, we find a strong and statistically significant correlation between the millimeter luminosity $L_{\rm mm}$ and the gamma-ray luminosity $L_\gamma$ across multiple ALMA bands and blazar subclasses.
A partial correlation analysis following Padovani (1992) demonstrates that this correlation remains significant after removing the redshift dependence, and its persistence despite non-simultaneous observations suggests that millimeter emission traces the time-averaged jet power underlying the gamma-ray emission.

\item
When divided into blazar subclasses, the partial correlation coefficients
show a possible systematic trend along the blazar sequence. The
redshift-controlled correlation tends to be strongest for HSP BL~Lacs and
progressively weaker toward ISP BL~Lacs, LSP BL~Lacs, and FSRQs, although the
Fisher-transformed 95\% confidence intervals overlap substantially because of
the limited number of BL~Lac subclasses. This trend is therefore suggestive
rather than statistically conclusive. It is nevertheless consistent with a
picture in which the millimeter and GeV gamma-ray emission are more directly
connected in SSC-dominated HSP BL~Lacs, while in FSRQs the GeV emission is
more strongly affected by external Compton scattering, external photon fields,
and Compton dominance.
This interpretation is further supported by the systematic variation of the Fermi-LAT photon index, which becomes progressively harder from FSRQs to HSP BL~Lac objects, consistent with the transition from EC-dominated to SSC-dominated high-energy emission along the blazar sequence.
\end{enumerate}

Overall, these results indicate that millimeter emission provides a useful,
time-stable tracer of the jet power underlying the high-energy emission, while
the strength of the millimeter--gamma-ray connection may depend on the
dominant inverse-Compton emission mechanism along the blazar sequence.
They also demonstrate the scientific potential of the ALMA calibrator archive as a unique statistical laboratory for blazar studies and provide a foundation for future population studies of millimeter-selected blazars.

\begin{acknowledgments}
We would like to acknowledge Francesco Massaro for helpful comments and suggestions.
This work was supported by JST SPRING, Grant Number JPMJSP2108.
K.N. was supported by the International Graduate Program for Excellence in Earth-Space Science (IGPEES), the University of Tokyo, and by JSPS Research Fellowship for Young Scientists (Grant Number JP26KJ0864). 
Y.I. is supported by JST FOREST Program (Grant Number JPMJFR256L, Japan) and
JSPS KAKENHI grant Nos. JP22K18277, JP26H00604, and JP26K22331.
This paper uses the ALMA data: ADS/JAO.ALMA\#2011.0.00001.CAL.
ALMA is a partnership of ESO (representing its member states), NSF (USA), and NINS (Japan), together with NRC (Canada), MOST and ASIAA (Taiwan), and KASI (Republic of Korea), in cooperation with the Republic of Chile. The Joint ALMA Observatory is operated by ESO, AUI/NRAO and NAOJ. The National Radio Astronomy Observatory is a facility of the National Science
Foundation operated under cooperative agreement by Associated Universities, Inc.
Data analysis was conducted using the Multiwavelength Data Analysis System, operated by the Astronomy Data Center (ADC) at the National Astronomical Observatory of Japan. 
\end{acknowledgments}


\bibliography{sample701}{}
\bibliographystyle{aasjournalv7}



\end{document}